\documentstyle[12pt]{article}        % Simple LaTeX, use with DINA4 as follows
\input epsf
\newlength{\dinwidth}
\newlength{\dinmargin}
\def\lsim{\mathrel{\rlap{\lower4pt\hbox{\hskip1pt$\sim$}}
  \raise1pt\hbox{$<$}}}	  %less than or approx. symbol
\def\gsim{\mathrel{\rlap{\lower4pt\hbox{\hskip1pt$\sim$}}
  \raise1pt\hbox{$>$}}}	  %greater than or approx. symbol

\begin{document}
\vspace*{1cm} 
\begin{center} 
\begin{Large} 
\begin{bf} 
Vector Meson
production in $ep \rightarrow epV$\\ 
\end{bf} 
\end{Large} 
\vspace*{5mm}
\begin{large} 
W. Koepf$^a$, P. V. Landshoff$^b$, E. M. Levin$^{c,d}$, N. N. Nikolaev$^{e,f}$
\\ 
\end{large} 
\end{center} % MM Change addresses as appropriate 
$^a$ Department of Physics, The Ohio State University, Columbus, OH 43210, 
     USA\\ 
$^b$ DAMPT, University of Cambridge, Cambridge CB3 9EW, UK\\ 
$^c$ Physics Division, Argonne National Laboratory, Argonne, IL60439, USA\\ 
$^d$ Theory Department, Petersburg Nuclear Physics Institute, 188350 Gatchina\\
$^{~\,}$ St. Petersburg, Russia\\ 
$^e$  ITKP, Universit\"at Bonn, Nu{\ss}allee 14-16, Bonn D-53115, FRG\\
$^f$  L.D.Landau Inst. Theor. Physics, Kosygina 2, 117334 Moscow, Russia\\
\begin{quotation} 
\noindent {\bf Abstract:} The
diffractive production of vector mesons in $ep$ interactions at low
$x$ is a subject of heated debates. This chapter consists of four
contributions written by the original authors and expresses the
possible scenarios which are to be investigated experimentally.
\end{quotation}

\section{Hard diffractive vector meson production}
\centerline{contributed by {\bf L. Frankfurt, W. Koepf and M. Strikman}}

The derivation of our QCD formulas \cite{Brod94} consists of three essential
steps: 
\hfill \break 
(i) The process factorizes into three stages: 
the creation of a quark-gluon wave packet, the interaction of this
packet with the target, and the formation of the vector meson.
The wave packet's large coherence length, ${1\over 2m_N x}$, justifies 
using completeness over the intermediate states.
\hfill \break 
(ii) For longitudinal polarization, the end point contribution is suppressed 
by $1/Q^2$, which supports applying the factorization theorem.
This important result follows from the well known light-cone
wave function of a photon and conformal symmetry of pQCD, which dictates the 
dependence of the vector meson's asymptotic wave function on the momentum 
fraction carried by the quarks. 
For transverse polarizations, the onset of the hard regime is expected at 
much larger $Q^2$ since large distance effects are suppressed only by the 
square of a Sudakov type form factor, $exp(-{4\alpha_s\over 3\pi} 
\ln^2 {Q^2\over k_t^2})$. 
\hfill \break 
(iii) As a result of the QCD factorization theorem, the hard amplitude 
factorizes from the softer blob \cite{CFS}.  Thus, within the leading 
$\alpha_s\ln Q^2$ approximation, the cross section of hard diffractive 
processes can be written in terms of the distribution of bare quarks within 
the vector meson and the gluon distribution in the target.

The respective cross section can thus be expressed through the 
light-cone wave function of the vector meson, $\psi_V(z,b=0)$, a well defined 
and intensively researched object in QCD.  In addition, there is not much 
freedom in the choice of this wave function since the absolute normalization
of the cross section is related to the vector meson's leptonic decay width, 
$\Gamma_{V\rightarrow e^+e^-}$.  Our numerical analysis has found a number of 
distinctive features of these hard diffractive processes (see Ref.\cite{FKS} 
and references therein):

\begin{enumerate}

\item A significant probability of small transverse size ($b_{q\bar q} 
\approx 3/Q$) minimal Fock $q\bar q$ configurations within the vector meson's 
light-cone wave function.

\vspace{-0.10cm}

\item A fast increase of the cross section at small $x$ and a
relatively slow $Q^2$ dependence, both resulting from the $Q^2$ evolution of
the parton distributions. 

\vspace{-0.10cm}

\item To avoid contradiction with $b$-space unitarity,
the increase of the cross section with decreasing $x$ should slow down.
For $Q^2 \sim 5$ GeV$^2$, this is expected at $x \sim 10^{-3}$ \cite{FKS}.

\vspace{-0.10cm}

\item For longitudinal polarization, an almost flavor and energy independent 
$t$-slope, while for transverse polarizations, soft QCD may reveal itself
in a larger value as well as an energy dependence of the latter.

\vspace{-0.10cm}

\item At large $Q^2$, the diffractive electroproduction ratio of $\rho$
and $\phi$ mesons follows from the form of the e.m.~current in the standard
model, i.e.~restoration of $SU(3)$ symmetry.  Some enhancement of the relative 
yield of the $\phi$ meson is expected due to its smaller size.  

\vspace{-0.10cm}

\item The diffractive photoproduction of $J/\psi$ mesons is dominated by 
relativistic $c\bar c$ configurations. Significant diffractive photoproduction 
of $\Upsilon$ mesons.

\vspace{-0.10cm}

\item Large cross sections for the production of excited states, 
$ep \rightarrow epV'$, with a ratio proportional to 
$M_{V'}\,\Gamma_{V'\rightarrow e^+e^-}$, in qualitative
difference from photoproduction processes.

\vspace{-0.10cm}

\item Model estimates found large $1/Q^2$ corrections to the basic 
formulas resulting from quark Fermi motion within the vector meson and 
from shadowing effects evaluated within the eikonal approximation. 
However, the reliability of these estimates is still unclear since
similar corrections follow from the admixture of $q\bar qg$ 
components to the vector meson's wave function and because the elastic eikonal
approximation is inappropriate.  Note that, in an exact calculation, the
contribution of more than two rescatterings by the $q\bar q$ pair should be 
zero due to energy-momentum conservation.

\end{enumerate}

\section{$\gamma^*$ p $\to$ V p {\bf at small} $t$}
\centerline{contributed by {\bf P. V. Landshoff}}

All models~\cite{don1,don1a,cud,lag,ginz,dos}
couple the $\gamma ^*$ to the vector meson $V$ through a
simple quark loop, to which is attached a pair of gluons which
interact with the proton. The models differ in two essentials: what
they assume about the wave function that couples $V$ to the $q\bar q$,
and what they assume about how the gluons interact with the proton.

Because the models have the same quark loop, there is general agreement that
the $\gamma ^*$ and the $V$ should have the same helicity,  and that
$$
{\sigma _L\over\sigma _T}\propto {Q^2\over m_V^2}
\eqno(1)
$$
so that at large $Q^2$ longitudinal production dominates. Presumably the
detailed dependence of ${\sigma _L/\sigma _T}$ on $Q^2$ is sensitive
to the form of the wave function. The very simple form assumed by DL fits
the data reasonably well ~\cite{don2}, 
but more theoretical work is needed to decide
just how much can be learnt about the wave function from this. Also,
according to (1), for heavier flavours it will need a larger $Q^2$ to
achieve dominance of the longitudinal polarisation; it is likely that 
most of the $J/\psi$ production at HERA will be transverse. 

The way in which the gluons couple to the proton will determine the
$W$-dependence. If they couple through a soft pomeron, then
$$
{d\sigma \over dt}=f(t,Q^2)\,e^{4(\epsilon -\alpha '|t|)\log W}
\eqno(2)
$$
with $\epsilon\approx 0.08$ and $\alpha '=0.25$ GeV$^{-2}$.
If the data find a larger value of $\epsilon$, this
may be a sign of the BFKL pomeron (though this is unlikely ~\cite{don3}), 
or of whatever other mechanism is responsible for the rapid rise seen in 
$\nu W_2$ at small $x$. Some of the models seek to make a direct connection
between the energy dependence of exclusive vector electroproduction and
the rise of $\nu W_2$ at small $x$, but there are theoretical problems
in this. The soft-pomeron form (2) predicts that the $t$-slope changes
with $W$ in a particular way; if the soft pomeron is not involved the
forward-peak shrinkage will surely not ocurr at the same rate and
is likely to be significantly slower, though this is not understood.
Notice that $f(t,Q^2)$ contains the square of the elastic form factor of
the proton and so is not a simple exponential: the $t$-slope will vary
with $t$. Notice also that its measurement is particularly sensitive
to any contamination from nonelastic events, which become increasingly
important as $|t|$ increases.

\section{Shadowing corrections in diffractive QCD leptoproduction 
of vector mesons}
\centerline{contributed by {\bf E. Gotsman, E. Levin and U. Maor}}

In our paper \cite{GLMPSI} the formulas for the shadowing corrections
( SC ) for vector meson diffractive dissociation ( DD ) in DIS have
been obtained within the framework of the DGLAP evolution equation in
the region of low $x$.  It is shown that the rescatterings of the
quarks is concentrated at small distances ( $ r^2_{\perp}
\,\propto\,\frac{1}{Q^2z(1 - z) + m^2_q}$ ) and can be treated
theoretically on the basis of perturbative QCD.

The numerical calculation of the damping factor defined as:
\begin{equation} \label{1PSI}
D^2\,\,=\,\,\frac{ \frac{d \sigma( \gamma^* p \rightarrow Vp)}{d t}
\,[\,with \,\,\,SC\,]}{ \frac{d \sigma( \gamma^* p \rightarrow Vp)}{d t}
\,[\,without \,\,\,SC\,]}\,|_{t = 0}
\end{equation}
shows that the SC (i) should be taken into account even at HERA
kinematic region and they generate the damping factor of the order of
0.5 for $J/\Psi$ production at $W = 100 - 200 GeV$ for $Q^2 = 0 - 6
GeV^2 $ ( see ~\cite{GLMPSI} for details); (ii) the value of the
SC is bigger than uncertainties related to the unknown nonperturbative
part of our calculations, and (iii) DD in vector meson for DIS can be
used as a laboratory for investigation of the SC.

The calculation of the SC for the gluon structure function depends on
a wide range of distances including large ones
($r^2_{\perp}\,>\,\frac{1}{Q^2z(1 - z) + m^2_q}$).  This causes a
large uncertainty in the pQCD calculations which, however, become
smaller at low $x$. We show that the gluon shadowing generates
damping, which is smaller or compatible with the damping found in
the quark sector.

The cross section of the vector meson DD is shown to be proportional
to $( \frac{d F_2^{exp}}{ d \ln Q^2} )^2 $ ~\cite{GLMPSI}.
This formula takes into account all possible SC and it is derived
in the leading log approximation of pQCD (for the GLAP evolution).  It
means that the experimental data for DD for vector meson production
provide information about $d F_2 / \ln Q^2 $, from which we could
extract the gluon structure function in the DGLAP evolution equation
in the region of low $x$.

\section{Vector mesons}
\centerline{contributed by {\bf N. N. Nikolaev and B. G. Zakharov}}

The amplitude of exclusive vector meson production ~\cite{kopel1,kopel2,
nem1,nem2,nikol} 
reads (we suppress the polarization subscripts T and L)
${\cal M} = \int d^{2}\vec{r}dz
\Psi^{*}_{V}(z,\vec{r})\sigma(x,r)\Psi_{\gamma^{*}}(z,\vec{r})$,
where $\Psi_{V,\gamma^{*}}$ are the color dipole distribution
amplitudes and $\sigma(x,r)$ is the color dipole cross section.  On
top of the gBFKL component which dominates $\sigma(x,r)$ at $r\lsim
R_{c}=0.3$\,fm, at larger $r$ in $\sigma(x,r)$ there is a soft
component. The non-negotiable prediction is that at a sufficiently
small $x$ the rising gBFKL component takes over at all $r$. The small
but rising gBFKL contribution provides a viable description of the
rise of soft cross sections.  ${\cal M}_{T,L}$ are dominated by $r
\approx r_{S}=6/\sqrt{Q^{2}+ m_{V}^{2}}$. The large value of the
scanning radius $r_{S}$ is non-negotiable and makes vector meson
production at best semiperturbative, unless $Q^{2}+m_{V}^{2} \gsim
$ 20-40\,GeV$^{2}$, i.e., unless $r_{S}\lsim R_{c}$.  Because the
scanning radius $r_{S}$ is so large, the formulas for the production
amplitudes in terms of the vector wave function at the origin are of
limited applicability at the presently studied $Q^{2}$.  When $r_{S}
\lsim R_{c}$ and the soft contribution is small, one can relate ${\cal
M}_{T,L}$ to the gluon density in the proton but at a very low
factorization scale $q_{T,L}^{2} = \tau (Q^{2}+m_{V}^{2})$ with $\tau
=$0.05-0.2 depending on the vector meson.  The energy dependence of
vector meson production at $r_{S}
\sim 0.15$\,fm, i.e., $Q^{2}(\Upsilon)\sim 0$ and
$Q^{2}(J/\Psi) \sim 100$\,GeV$^{2}$ and $Q^{2}(\rho)\sim
200$\,GeV$^{2}$ probes the asymptotic intercept of the gBFKL pomeron.
The major gBFKL predictions are:
\begin{enumerate}
\item
A steady decrease of $R_{LT}$ with $Q^{2}$ in $\sigma_{L}/\sigma_{T} =
R_{LT}Q^{2}/m_{V}^{2}$ with $R_{LT}\sim 1$.
\item
When fitted to $W^{4\Delta}$, the effective intercept $\Delta$ is
predicted to rise with $Q^{2}$. It also rises with $W$ and flattens at
a $Q^{2}$ independent $\Delta \approx 0.4$ at a very large $W$. The
universal energy dependence is predicted for all vector mesons if one
compares cross sections at identical $\bar{Q}^{2}$.
\item
Comparing the $Q^{2}$ dependence of ${\cal M}_{T,L}$ makes
no sense, the real parameter is $r_{S}$ and/or $\bar{Q}^{2}=
Q^{2}+m_{V}^{2}$, the ratios like $(J/\Psi)/\rho$ exhibit
wild $Q^{2}$ dependence but we predict the flavor dependence
disappears and these ratios are essentially flat vs. $\bar{Q}^{2}$.
The $Q^{2}$ dependence must follow the law $\propto \bar{Q}^{2n}$,
where $n$ is about flavor independent, typically $n\sim -2.2$
at HERA. The $\propto \bar{Q}^{2n}$ fits are strongly recommended.
\item 
Strong suppression of the $2S/1S$ cross section ratios by the node
effect is a non-negotiable prediction, these ratios are predicted to
rise steeply and then level off on a scale $Q^{2} \lsim
m_{V}^{2}$. Steady rise of these ratios with energy is predicted.  For
the D-wave vector mesons the ratio $D/1S$ is pedicted to have a weak
$Q^{2}$ dependence in contrast to the $2S/1S$ ratio.
\item
The gBFKL pomeron is a moving singularity and we predict the
conventional Regge shrinkage of the diffraction cone. The rise of
the diffraction slope by $\sim 1.5$\, GeV$^{-2}$, which is about universal
for all vector mesons and at all $Q^{2}$, is predicted to take place
from the CERN/FNAL to HERA energies.  An inequality of diffraction
slopes $B(2S) < B(1S)$ is predited. For the $\rho'(2S)$ and
$\phi'(2S)$ the diffraction cone can have a dip and/or flattening at
$t=0$.  For the 1S states, the diffraction cone for different vector
mesons must be equal if compared at the same $\bar{Q}^{2}$.
\end{enumerate}

\end{document}